# Field-induced magnetic charge in a cubic Laves compound UAl$_2$

S. W. Lovesey[1,2*], D. D. Khalyavin[1] and G. van der Laan[2]

[1]ISIS Facility, STFC, Didcot, Oxfordshire OX11 0QX, UK

[2]Diamond Light Source, Didcot, Oxfordshire OX11 0DE, UK

**Abstract**. Magnetic diffraction of polarized neutrons by the cubic Laves compound UAl$_2$ in a magnetic field has unveiled weak Bragg spots that are nominally forbidden. On the one hand, they can be viewed as magnetic analogues of the basis-forbidden (2, 2, 2) reflection in diamond-type structures that has been painstakingly and frequently investigated over almost a century. Alternatively, the pattern of weak intensities can be assigned to Dirac multipoles imbedded in field-induced magnetic charge. To this end, a published diffraction pattern is successfully confronted with intensities calculated from the appropriate magnetic space-group (Imm′a′) that includes Dirac dipoles (anapoles) to describe the basis-forbidden magnetic reflections (H$_o$ + K$_o$ + L$_o$ = 4n + 2), and conventional (axial) dipole and octupole multipoles to describe basis-allowed magnetic reflections.

## 1. Introduction

Using the conventional approximation of spherical ions in elemental materials that assume a diamond-type structure, Bragg reflections whose Miller indices sum up to 4n + 2 (H$_o$ + K$_o$ + L$_o$ = 4n + 2, n an integer) are forbidden. X-ray diffraction measurements of these basis-forbidden Bragg spots in diamond structure materials have been performed since 1921 [1]. Intensity in the weak reflections is created by a difference between the actual electronic charge distribution and a space-group allowed array of spheres. Moreover, the two ions in the basis of diamond-type structures differ by a point inversion and the weak intensity arises from an admixture of valence electrons with opposing parities. In consequence, intensities of basis-forbidden Bragg spots give valuable information about the valence electron distribution and they have been painstakingly investigated by neutron and x-ray diffraction; see, for example, studies of Ge and Si in references [2, 3].

Compounds with the C15 cubic Laves structure possess a similar class of forbidden reflections. Specifically, the rare earth and actinide compounds SmAl$_2$ [4] and UAl$_2$ [5] display reflections indexed by H$_o$ + K$_o$ + L$_o$ = 4n + 2 that are nominally due to Al nuclei or Al electron distributions alone. The same condition applies to Bragg diffraction when the Laves structure possesses ferromagnetic order, either induced by an applied magnetic field or a spontaneous development below a phase transition. In the case of magnetic diffraction, magnetic charge is responsible for intensities of the forbidden reflections. The corresponding unit-cell structure factors for magnetic diffraction can be expressed with electronic multipoles that are both parity-odd (polar) and time-odd (magnetic), which are labelled Dirac multipoles because the elementary Dirac monopole - yet to be observed - possesses identical discrete symmetries. In this communication we interpret intensities of Bragg spots from the intermetallic alloy UAl$_2$ that are induced by an applied magnetic field [5]. Intensities were measured by neutron diffraction enhanced by polarization analysis that assures the observed signal has a magnetic origin, to a very good approximation. Allowed Bragg spots are here assigned to U $5f^3$ electrons, whereas Dirac multipoles observed at forbidden reflections are nominally assigned to a configuration $5f^3$ - $6d^1$ (a distribution of Al electrons at the U site can represented by U 6d states). Dirac multipoles have been investigated in numerous materials using the technique of resonance-enhanced x-ray Bragg diffraction [6, 7, 8].

The operator for magnetic neutron scattering **Q** is time-odd and an axial dipole [9]. Its expectation value ⟨**Q**⟩, which describes Bragg diffraction, possesses a sign determined by the magnetic field that changes when the polarity of the field is reversed. The operator **Q** is composed of electronic variables (operators) for position (**R**$_j$), spin (**s**$_j$) and linear momentum (**p**$_j$) that form multipole operators of integer rank $K$ that may be grouped according to their spatial parity.

Parity-even multipoles ⟨**T**$^K$⟩ are constructed from matrix elements ⟨σ$l$|**T**$^K$|σ′$l$′⟩ in which electronic angular momentum obeys $l + l'$ even, and σ denotes a spin projection. The rank $K$ can be even or odd, and the maximum $K = (2l + 1)$ for an atomic shell ($l = l'$). Multipoles with $K$ even are allowed when electrons require more than one J-manifold of states. The dipole ⟨**T**$^1$⟩ possesses a useful approximation in terms of total angular momentum **L** and spin **S**. For very small deflections of the neutron beam ⟨**T**$^1$⟩ = ⟨**L** + 2**S**⟩/2, which is a result first derived by Schwinger (1937) [10].

Dirac multipoles are constructed from matrix elements ⟨σ$l$|**O**$^K$|σ′$l$′⟩ in which electronic angular momentum obeys $l + l'$ odd. In the expectation value ⟨**Q**⟩ these multipoles combine with the neutron wavevector, **k**, to form an axial vector that is time-odd [9]. Examples of Dirac dipoles in ⟨**Q**⟩ include a spin anapole ⟨(**S** × **R**)⟩ and an orbital anapole **Ω** = ⟨(**L** × **R**) − (**R** × **L**)⟩ (expressions for multipoles formed with **L** and **R** are more complicated because the operators do not commute).

To date, Dirac multipoles provide a comprehensive interpretation of the diffraction pattern of polarized neutrons obtained from the pseudo-gap phase of the ceramic superconductor Hg1201 [11]. The magnetic symmetry of the phase has a subtlety, according to us [12, 13]. Copper ions in the paramagnetic phase use sites with a centre of spatial inversion symmetry, which forbids parity-odd Cu multipoles. However, the inversion symmetry is replaced by anti-inversion $\bar{1}'$ in the magnetic phase, which forbids axial magnetism and concomitant signals in NMR, NQR and muon spin-rotation experiments. The phase of Hg1210 investigated by Bourges *et al.* [11] emerges as pure magnetic charge in our scenario. It is predicted to display an interesting optical property, namely, the Kerr effect that is traditionally associated with ferromagnetism [12, 14]. Confidence in our scenario for polarized neutron diffraction by the pseudo-gap phase of Hg1201 is bolstered by a subsequent calculation of its electronic structure and Dirac multipoles [15].

## 2. Crystal structure

$UAl_2$ assumes the chemical structure Fd$\bar{3}$m, (#227, C15 cubic Laves structure) with a cell length $a_o \approx 7.78$ Å. The face-centred $Cu_2Mg$-type structure is depicted in Figure 1. Uranium ions are in sites 8a in #227 with site symmetry $\bar{4}3m$ ($T_d$) and an origin (1/8, 1/8, 1/8). Corresponding Miller indices are denoted ($H_o$, $K_o$, $L_o$), with conditions $H_o + K_o$, $H_o + L_o$ & $K_o + L_o$ even demanded by F-centring. The compound is paramagnetic.

## 3. Magnetic properties

Experimental results for polarized neutron diffraction by $UAl_2$ are available for magnetization induced parallel to the crystal axis [1, 1, 0]. The applied field had strength of 4.25 T and the sample was held at a temperature of 4.2 K [5]. The corresponding magnetic crystal-class $D_{2h}$ ($C_{2h}$) = mm'm' contains a centre of inversion symmetry, and a non-linear magnetoelectric effect is allowed.

Unit-cell diffraction amplitudes are derived from an electronic structure factor $\Psi_{K,Q}$ that respects all elements of translation symmetry and point symmetry in orthorhombic Imm'a' (#74.559). It is a linear combination of multipoles $\langle U^K_Q \rangle$, representing aforementioned $\langle T^K_Q \rangle$ or $\langle O^K_Q \rangle$, that encapsulate spin and orbital degrees of freedom in the electronic ground-state, with $\langle ... \rangle$ an expectation value, or time-average, of the enclosed spherical tensor-operator. The integer rank $K$ and projections $Q$ obey $-K \leq Q \leq K$. Multipoles are Hermitian. The complex conjugate of an Hermitian multipole satisfies $\langle U^K_Q \rangle^* = (-1)^Q \langle U^K_{-Q} \rangle$, and we choose the phase convention $\langle U^K_Q \rangle = \{\langle U^K_Q \rangle' + i \langle U^K_Q \rangle''\}$ for real $\langle U^K_Q \rangle'$ and imaginary $\langle U^K_Q \rangle''$ components.

Signatures $\sigma_\pi = \pm 1$ and $\sigma_\theta = -1$ (magnetic) denote discrete symmetries of parity and time of $\langle U^K_Q \rangle$, with $\sigma_\pi \sigma_\theta = -1$ for parity-even multipoles, and $\sigma_\pi \sigma_\theta = +1$ for Dirac multipoles.

*3.1. Neutron diffraction*

The amplitude for magnetic Bragg diffraction is $\langle \mathbf{Q}_\perp \rangle = [\boldsymbol{\kappa} \times (\langle \mathbf{Q} \rangle \times \boldsymbol{\kappa})]$ with a unit vector $\boldsymbol{\kappa} = \mathbf{k}(H_o, K_o, L_o)/k$. The intermediate operator in $\mathbf{Q}_\perp$ can be written [9],

$$\mathbf{Q} = \exp(i\mathbf{R}_j \cdot \mathbf{k}) [\mathbf{s}_j - (i/\hbar k)(\boldsymbol{\kappa} \times \mathbf{p}_j)]. \qquad (1)$$

The implied sum on $j$ includes unpaired electrons, because matrix elements of paired electrons are zero. Note that $\mathbf{Q}$ is arbitrary to within any function proportional to $\boldsymbol{\kappa}$.

We denote components of $\langle \mathbf{Q}_\perp \rangle$ formed with axial and polar electronic operators by $\langle \mathbf{Q}_\perp \rangle^{(\pm)}$ with $\langle \mathbf{Q}_\perp \rangle = \langle \mathbf{Q}_\perp \rangle^{(+)} + \langle \mathbf{Q}_\perp \rangle^{(-)}$. Multipoles in $\langle \mathbf{Q}_\perp \rangle^{(+)}$ have ranks $K = 1$ through $K = 7$. Radial integrals diminish in magnitude with increasing $K$ [16], and an interpretation of good quality is obtained here with a dipole and two octupoles allowed by magnetic symmetry. Quadrupoles are set aside because they are zero for a uranium ground-state derived from a manifold in $5f^3$ (our explorations of the Bragg diffraction patterns demonstrate no differences between $5f^2$ and $5f^3$, to a good approximation [9]). For the weak reflections attributed to $\langle \mathbf{Q}_\perp \rangle^{(-)}$ we use allowed multipoles of the lowest rank, which are anapoles (Dirac dipoles) when magnetization is parallel to [1, 1, 0].

*3.2. Unit-cell structure factors*

Results are calculated from the magnetic space-group Imm'a' [17]. Uranium ions are in sites 4e at an origin (0, 1/4, 1/8) and possess symmetry mm'2'. (In polar site symmetry mm2 the position of an U ion along the c-axis is not fixed, in contrast to the restriction imposed by $\bar{4}$3m (non-polar) site symmetry in the parent structure.) A basis {(1/2, 1/2, 0), (1/2, −1/2, 0), (0, 0, −1)} means that Miller indices satisfy,

$$h = (H_o + K_o)/2, \; k = (H_o - K_o)/2, \text{ and } l = -L_o. \tag{2}$$

Miller indices $h$ & $k$ are integers by virtue of the F-centring condition in the cubic Laves structure. Cell lengths are $a = b = a_o/\sqrt{2}$ & $c = a_o$. The right-handed orthonormal basis defines local principal-axes ($\xi, \eta, \zeta$) with $\boldsymbol{\xi} = (1, 1, 0)/\sqrt{2}$, $\boldsymbol{\eta} = (1, -1, 0)/\sqrt{2}$ and $\boldsymbol{\zeta} = (0, 0, -1)$, and the magnetic cell is depicted in Figure 1.

Uranium ions in UAl$_2$ are described by an electronic structure factor,

$$\Psi_{K,Q}(\text{Imm'a'}) = 2 \exp(i\pi\mathcal{W}/4) \langle U^K{}_Q \rangle [1 + \sigma_\pi \exp(-i\pi\mathcal{W}/2)], \tag{3}$$

with $\mathcal{W} = (H_o - K_o - L_o)$ and $(H_o + L_o)$ even. Several features of this expression merit comment. First, $\Psi_{K,Q}(\text{Imm'a'})$ is proportional to the uranium multipole that is bound by constraints imposed by site symmetry. Second, the time signature is not explicitly present in the electronic structure factor, but it does impact the U multipole (with $\sigma_\theta = -1$ for $\langle \mathbf{Q}_\perp \rangle$). Also, $\Psi_{K,Q}(\text{Imm'a'}) = 0$ for $\sigma_\pi = +1$ and $\exp(-i\pi\mathcal{W}/2) \equiv \exp\{-i\pi(H_o + K_o + L_o)/2\} = -1$, where the equality is achieved by virtue of $(K_o + L_o)$ even. In other words, the absence condition for a diamond-type structure is preserved in the presence of ferromagnetic long-range order. Diffraction indexed by $\mathcal{W} = (4n + 2)$, where n is an integer, is due to Al nuclei that overlaps an electronic (magnetic) contribution from parity-odd multipoles ($\sigma_\pi = -1$). With this case, we have the result $\exp(i\pi\mathcal{W}/4) = i(-1)^n$ for the pre-factor in (3), and a change of sign in the expression upon a simultaneous change of sign to all Miller indices.

Site symmetry mm'2' [17] includes the constraint,

$$\langle U^K{}_Q \rangle = m_\eta' \langle U^K{}_Q \rangle = I \, \theta \, 2_\eta \langle U^K{}_Q \rangle = - \sigma_\pi \sigma_\theta (-1)^K \langle U^K{}_{-Q} \rangle. \tag{4}$$

Here, I and $\theta$ denote operators for spatial inversion and time reversal, while $2_\eta$ is a diad rotation operator on the $\eta$-axis depicted in Figure 1. Identity (4) uses the condition $Q$ odd imposed by $2_\zeta' = (\theta \, 2_\zeta)$. Parity-even dipoles are restricted to the $\xi$-axis, by construction, while anapoles lie along the $\eta$-axis, and the two motifs are displayed in Figure 1.

**4. Confrontation with diffraction pattern**

The neutron polarization technique (flipping ratio) employed by the authors gives access to the real part of the component of $\langle \mathbf{Q}_\perp \rangle$ parallel to the applied magnetic field [4, 9], and this component equates to $\langle \mathbf{Q}_{\perp,\xi} \rangle$ with our principal-axes ($\xi$, $\eta$, $\zeta$) in Figure 1. Experimental results are presented as a form factor for the uranium ion that we denote by f($h$, $k$, $l$). Angular anisotropy in the diffraction pattern is expressed in terms Cartesian components of the Bragg wavevector,

$$\kappa_\xi = h\sqrt{(2/[\ H_o^2 + K_o^2 + L_o^2])}, \quad \kappa_\eta = k\sqrt{(2/[\ H_o^2 + K_o^2 + L_o^2])},$$

$$\kappa_\zeta = - L_o/\sqrt{[\ H_o^2 + K_o^2 + L_o^2]}, \qquad (5)$$

with $(\kappa_\xi)^2 + (\kappa_\eta)^2 + (\kappa_\zeta)^2 = 1$. We first consider Bragg spots in the observed pattern that are allowed by the structure factor (3) evaluated for axial magnetism. A good result for an interpretation of these data will add confidence to an interpretation thereafter of weak reflections due to Dirac multipoles.

*4.1. Basis-allowed reflections*

A value for $f^{(+)}(h, k, l)$ is derived from a universal expression for $\langle \mathbf{Q}_\perp \rangle^{(+)}$ [9],

$$f^{(+)}(h, k, l) \approx 35.2 \,[\langle j_0 \rangle + q \,\langle j_2 \rangle + \{\langle j_2 \rangle + p \,\langle j_4 \rangle\}\{r\,[1 - 5\kappa_\zeta^2 + \kappa_\xi^2 (15\kappa_\zeta^2 - 1)]$$

$$+ t\,[\kappa_\xi^2 (1 - \kappa_\xi^2) + \kappa_\eta^2 (3\kappa_\xi^2 - 1)]\}/\{2(1 - \kappa_\xi^2)\}], \qquad (6)$$

where the pre-factor agrees with the reported induced magnetic moment (two values for the pre-factor appear in the paper (34.4 in the text and 35.2 in the figure caption) and the small difference does not change the robustness of the fit displayed in Figure 3). We include in $f^{(+)}(h, k, l)$ allowed dipole and octupole moments set in principal axes ($\xi$, $\eta$, $\zeta$). In a fit to experimental data, radial integrals $\langle j_n \rangle$ with n = 0, 2 & 4 were calculated from interpolation formulae for the atomic configuration $5f^3$ ($U^{3+}$) for which J = 9/2 [16], and the principal component $\langle j_0 \rangle$ of $f^{(+)}(h, k, l)$ is illustrated in Figure 2. Table 1 and Figure 3 contain observed and calculated values of $f^{(+)}(h, k, l)$. Values for p, q, r & t inferred from diffraction data are p ≈ –2.75, q ≈ 1.65, r ≈ 0.35 and t ≈ –0.98.

We gain some insight on the physical meaning of these quantities from the following observations. For $5f^3$ ($^4$I) the saturation value of the parity-even dipole $\langle J, M|T^1_0|J, M \rangle$ = {(g/3) J [$\langle j_0 \rangle$ + (119/66) $\langle j_2 \rangle$]}, where the Landé g-factor g = 8/11 (J = M = 9/2). This exact result yields q = 119/66 ≈ 1.80, while q ≈ (2 − g)/g = 1.75 in the dipole approximation [9]. For the case in hand, and the same level of approximation, q ≈ $\langle L_\xi \rangle/\mu_o$, where $\mu_o$ is the field-induced magnetic moment. Quantities r and t are octupoles, with $\langle T^3_{+1} \rangle'$ = (2$\mu_o$ r/√21) {$\langle j_2 \rangle$ + p $\langle j_4 \rangle$} and ($\langle T^3_{+3} \rangle'/\langle T^3_{+1} \rangle'$) = − [t/(r√15)]. The experimental data imply that the two octupoles are of the same sign and similar in value with $\langle T^3_{+1} \rangle'/\langle T^3_{+3} \rangle' \approx 1.36$.

*4.2. Basis-forbidden reflections*

Results for a magnetic amplitude, or form factor, $f^{(-)}(h, k, l) = \langle \mathbf{Q}_{\perp,\xi} \rangle^{(-)}$ are reported for three weak reflections, namely, $f^{(-)}(2, 0, -2) = -0.7\,(2)$, $f^{(-)}(4, 2, -2) = -0.1\,(3)$ & $f^{(-)}(4, -2, -6) = +0.7\,(3)$ [5]. We attribute the basis-forbidden reflections to diffraction by anapoles depicted in Figure 1 and set aside higher-order Dirac multipoles. In this first approximation [9],

$$f^{(-)}(h, k, l) \approx |\kappa_\zeta| (-1)^n \,[i\,\langle \mathbf{n}_\eta \rangle\,(g_1) + 3\,\langle (\mathbf{S} \times \mathbf{n})_\eta \rangle\,(h_1) - \langle \mathbf{\Omega}_\eta \rangle\,(j_0)], \quad (7)$$

with $\mathcal{W} = (H_o - K_o - L_o) = (4n + 2)$. The three radial integrals $(g_1)$, $(h_1)$ & $(j_0)$ evaluated for $5f^3$ - $6d^1$ are displayed in Figure 2. The dipoles $\langle \mathbf{n}_\eta \rangle$, $\langle (\mathbf{S} \times \mathbf{n})_\eta \rangle$ & $\langle \mathbf{\Omega}_\eta \rangle$ are purely real, where $\mathbf{n}$ is an electronic (polar) vector $\mathbf{n} = \mathbf{R}/R$. The measured flipping ratio is proportional to the real part of $f^{(-)}(h, k, l)$ and consequently it contains no information on $\langle \mathbf{n}_\eta \rangle$. We infer from the experimental data that the spin and orbital anapoles are in a ratio of about 2:1, with $\langle (\mathbf{S} \times \mathbf{n})_\eta \rangle \approx 6.2$ and $\langle \mathbf{\Omega}_\eta \rangle \approx -3.4$. Errors on the estimates tell little of value, because the fractional errors on the weak intensities are quite large and could, most likely, be reduced in renewed experiments.

**5. Discussion**

Basis-forbidden reflections intense enough to be observed in the diffraction of polarized neutrons by field-polarized $UAl_2$ [5] are here assigned to Dirac multipoles imbedded in magnetic charge. Our arguments for the cubic Laves compound are informed by magnetic symmetry, and encompass a successful account of spatial anisotropy in the uranium wavefunction observed in 12 basis-allowed reflections [5]. Similar arguments have been successfully employed to infer the contribution of anapoles to Bragg diffraction patterns of $SmAl_2$ lightly doped with Gd [23], which is a ferromagnet below a temperature ≈ 127 K [4].

The narrow $5f_{5/2}$ band of $UAl_2$ contains states with more itinerant character, mainly U 6d and Al 3p. Occupations of 6d and 5f states in $UAl_2$ are almost equal, in fact, which is consistent with the presence of anapoles of a magnitude sufficient to be measured in diffraction. Band structure calculations for actinide Laves compounds show a strong variation of the ratio of 6d to 5f occupations, with values ≈ 0.87, 0.42 and 0.27 for $UAl_2$, $NpAl_2$ and $PuAl_2$, respectively, implying that the uranium compound under discussion is a favourable candidate material in which to study Dirac multipoles [22]. Evidence that magnetic polarization of Al ions is insignificant compared to U Dirac multipoles could be derived from a simulation of the electronic structure similar to the one performed for Hg1201 [15]. An estimate of Al polarization would be found in a map of the magnetization density constructed from a measured Bragg diffraction pattern [24]. However, the extent of the pattern and the precision required for a meaningful exercise is likely beyond practical realization.

When the magnetic field is applied along the [1, 1, 1] crystal axis of $UAl_2$ the diffraction pattern will be different from the one with magnetization induced along [1, 1, 0] discussed in the main text. SmAl$_2$ actually develops spontaneous ferromagnetism below a critical temperature ≈ 127 K with [1, 1, 1] the easy-axis [4]. The appropriate magnetic space-group for magnetization parallel to [1, 1, 1] is R$\bar{3}$m′ (#166.101, magnetic crystal-class $\bar{3}$m′, D$_{3d}$ (S$_6$)) in which U (or Sm) ions are in sites with symmetry 3m′., namely, 6c at an origin (0, 0, 7/8). A basis {(1/2, 0, −1/2), (0, −1/2, 1/2), (−1, −1, −1)} is hexagonal with cell lengths $a_h$ = $a_o/\sqrt{2}$ & $c_h$ = $a_o\sqrt{3}$. Orthonormal local principal-axes (ξ, η, ζ) in which multipoles are defined are derived from ξ = $\mathbf{a}_h/a_h$ = (1, 0, −1)/√2, η = ($\mathbf{a}_h + 2\mathbf{b}_h$)/($a_h\sqrt{3}$) ∝ $\mathbf{b}_h$* and ζ = − (1, 1, 1)/√3. An axial dipole (K = 1) parallel to the ζ-axis is allowed, by construction, while an anapole is strictly forbidden. Allowed Dirac multipoles include, ⟨O$^0_0$⟩, ⟨O$^2_0$⟩, ⟨O$^3_{+3}$⟩ = − ⟨O$^3_{-3}$⟩, ⟨O$^4_0$⟩ and ⟨O$^4_{\pm3}$⟩. (The Dirac monopole ⟨O$^0_0$⟩ ∝ ⟨**S** • **n**⟩ does not contribute to the amplitude for magnetic neutron scattering, although it is visible in light scattering [18, 19].) In which case, a first approximation to f$^{(-)}$(h, k, l) is made by Dirac quadrupoles, as is the case for the ceramic superconductor Hg1201 [12]. The radial integral for the Dirac quadrupole constructed from **S** and **n** is (h$_1$), which is illustrated in Figure 2 for U$^{3+}$ (5f$^3$ - 6d$^1$).

**Acknowledgements**

We have benefited from correspondence with Dr T. Chatterji (ILL). Reduced matrix-elements for a U ion were calculated by Professor Ewald Balcar (Vienna). We thank the referee for their diligence.

* Dedicated to the memory of Sir Roger Elliott who died May, 2018.

**Table 1.** Values of the form factor f$^{(+)}$ reported by Rakhecha *et al*. [5] with fractional errors, together with values calculated from (6) using p = −2.75, q = 1.65, r = 0.35 and t = −0.98. A goodness-of-fit R$_F$ = Σ |(f$^{(+)}$)$^2_{obs}$ − (f$^{(+)}$)$^2_{cal}$|/Σ (f$^{(+)}$)$^2_{obs}$ with a sum over 12 Bragg spots yields R$_F$ = 10.3%. A graphical display of the data is shown in Figure 3. Reflections are here labelled (H$_o$, K$_o$, L$_o$) as they appear in the original paper [5].

| Reflection | Observed | Calculated f$^{(+)}$ |
|---|---|---|
| (1, 1, 1) | 32.1 (7) | 34.1 |
| (0, 2, 2) | 29.0 (4) | 28.6 |
| (3, 1, 1) | 28.3 (5) | 26.0 |
| (0, 0, 4) | 20.9 (3) | 21.3 |
| (1, 3, 3) | 20.5 (3) | 22.4 |
| (4, 2, 2) | 20.3 (3) | 20.2 |
| (3, 3, 3) | 19.7 (4) | 19.0 |
| (5, 1, 1) | 19.6 (4) | 17.6 |
| (0, 4, 4) | 16.7 (4) | 16.2 |
| (4, 4, 4) | 9.0 (2) | 9.0 |
| (1, 5, 5) | 10.0 (4) | 10.5 |
| (7, 1, 1) | 10.9 (3) | 12.4 |

---------------------------------------------------------------------------------------------------

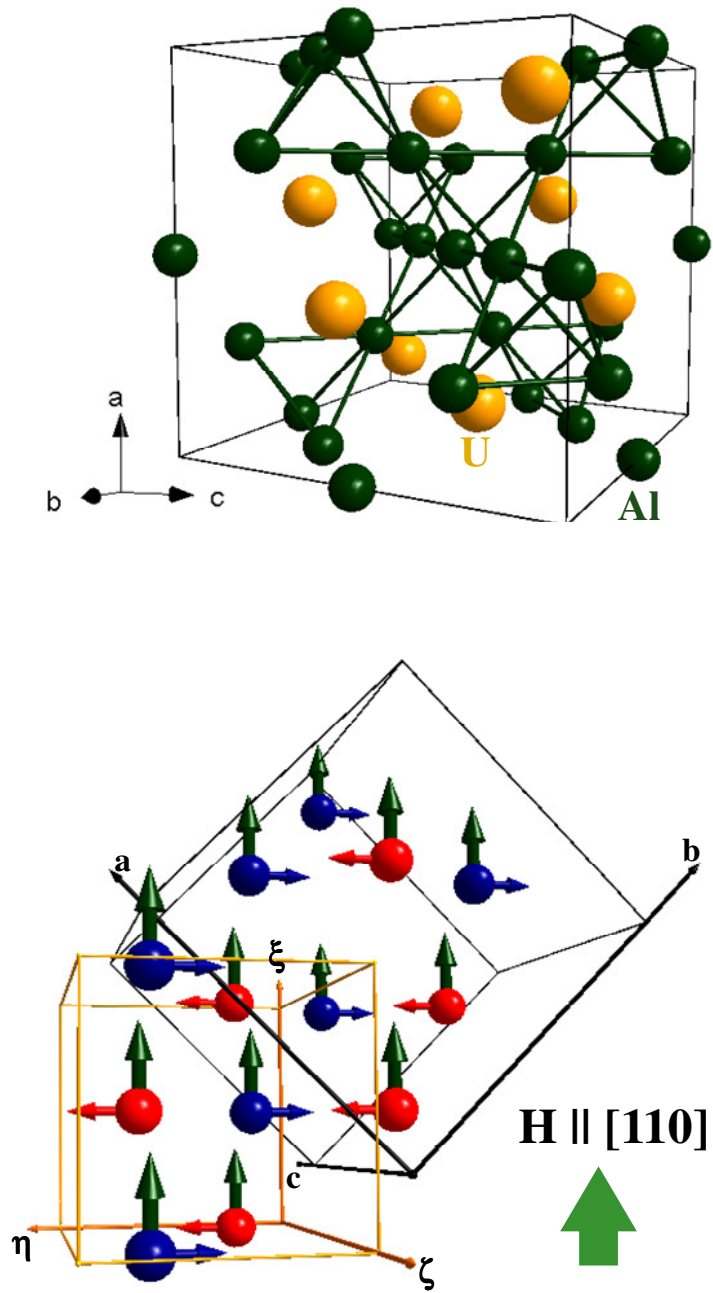

**Figure 1**. UAl$_2$: Top panel; crystal structure (#227, C15 cubic Laves) with U ions in yellow and Al ions in green, and cell edges (a, b, c). Bottom panel; magnetic dipoles induced on U ions by a magnetic field parallel to the crystal axis [1, 1, 0] (#74.559, Imm′a′). Cubic parent

cell outlined in black, and orthorhombic magnetic cell (ξ, η, ζ) with $\xi = (1, 1, 0)/\sqrt{2}$, $\eta = (1, -1, 0)/\sqrt{2}$ and $\zeta = (0, 0, -1)$ outlined in yellow. Green arrows are axial dipoles parallel to the ξ-axis, while blue and red arrows that lie along the η-axis denote anapoles related by point inversion.

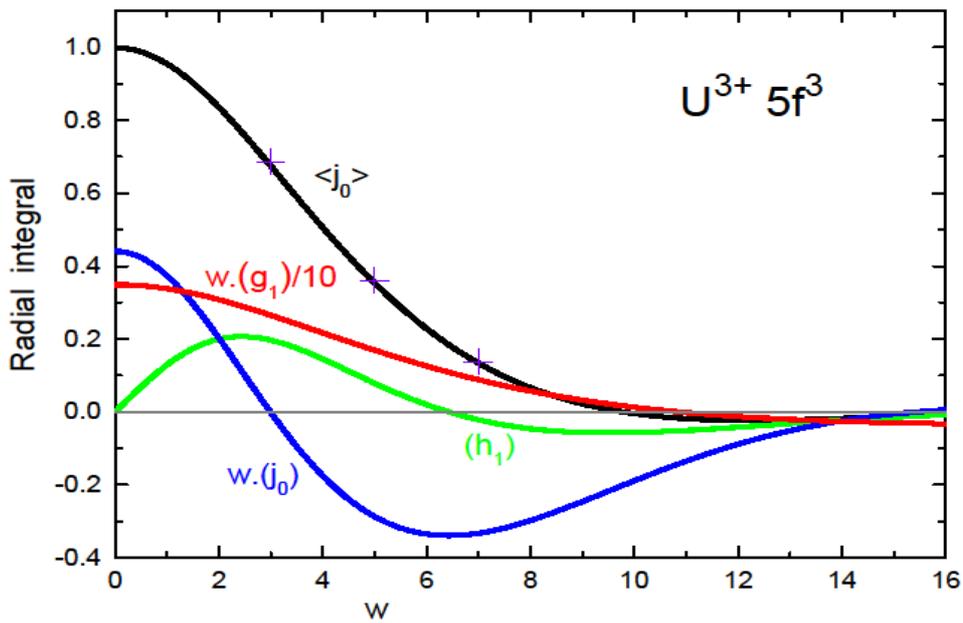

**Figure 2**. Radial integrals for anapoles in the form factor (7) are displayed as a function of a dimensionless variable $w = 12\pi a_o s$, where $a_o$ is the Bohr radius, while the standard variable for radial integrals $s$ is derived from the Bragg angle and neutron wavelength $s = \sin(\theta)/\lambda$. Legend: (—) $[w \times (g_1)]/10$, (—) $(h_1)$ & (—) $[w \times (j_0)]$. Note that $(g_1)$ and $(j_0)$ arise from the component of **Q** in (1) that contains the linear momentum operator and they are proportional to $1/w$ as the wavevector approaches zero. Atomic wavefunctions are $5f^3 - 6d^1$. (See, also, references [9, 20].) Also included in the figure is the standard radial integral $\langle j_0 \rangle$ that appears in the result (6) for $f^{(+)}(h, k, l)$. Results obtained with our $U^{3+}$ ($5f^3$) wavefunction are denoted by the continuous black curve, to which we added for comparison three values (+) derived from the standard interpolation formula [16].

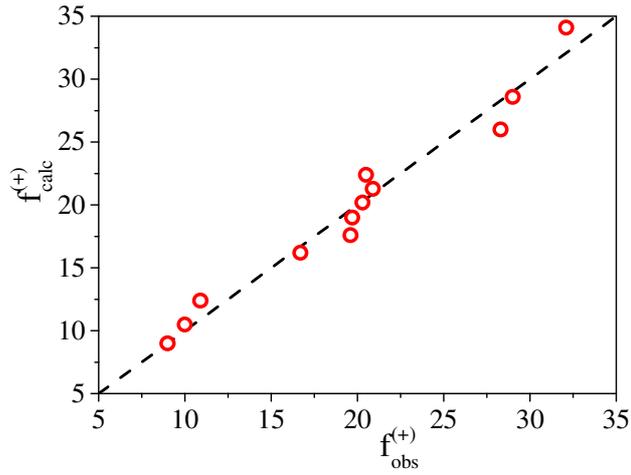

**Figure 3.** Observed and calculated form factors listed in Table 1 for basis-allowed reflections. Departures from the standard radial integral $\langle j_0 \rangle$ in the calculated amplitude (6) caused by angular anisotropy in the U wavefunction are adequately explained by axial octupoles alone, with axial triakontadipoles neglected. Note that even rank multipoles from the spin-orbital part of **Q** in (1) would contribute to the calculated form factor in the event that a U wavefunction possessed two or more manifolds, e.g., manifolds with $J = 9/2$ & $J = 11/2$.